www.nature.com/scientificreports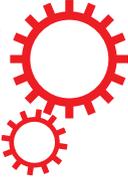

OPEN

Received: 22 March 2016
Accepted: 01 July 2016
Published: 17 August 2016# Targeted nanoconjugate co-delivering siRNA and tyrosine kinase inhibitor to KRAS mutant NSCLC dissociates GAB1-SHP2 post oncogene knockdown

R. Srikar[1], Dhananjay Suresh[2], Ajit Zambre[1], Kristen Taylor[3], Sarah Chapman[4], Matthew Leevy[4], Anandhi Upendran[5,6] & Raghuraman Kannan[1,2,7]A tri-block nanoparticle (TBN) comprising of an enzymatically cleavable porous gelatin nanocore encapsulated with gefitinib (tyrosine kinase inhibitor (TKI)) and surface functionalized with cetuximab-siRNA conjugate has been synthesized. Targeted delivery of siRNA to undruggable KRAS mutated non-small cell lung cancer cells would sensitize the cells to TKI drugs and offers an efficient therapy for treating cancer; however, efficient delivery of siRNA and releasing it in cytoplasm remains a major challenge. We have shown TBN can efficiently deliver siRNA to cytoplasm of KRAS mutant H23 Non-Small Cell Lung Cancer (NSCLC) cells for oncogene knockdown; subsequently, sensitizing it to TKI. In the absence of TKI, the nanoparticle showed minimal toxicity suggesting that the cells adapt a parallel GAB1 mediated survival pathway. In H23 cells, activated ERK results in phosphorylation of GAB1 on serine and threonine residues to form GAB1-p85 PI3K complex. In the absence of TKI, knocking down the oncogene dephosphorylated ERK, and negated the complex formation. This event led to tyrosine phosphorylation at Tyr627 domain of GAB1 that regulated EGFR signaling by recruiting SHP2. In the presence of TKI, GAB1-SHP2 dissociation occurs, leading to cell death. The outcome of this study provides a promising platform for treating NSCLC patients harboring KRAS mutation.NSCLC is diagnosed in an estimated 220,000 patients each year with five-year overall survival rates of 16 percent[1]. A recent report confirmed that 16 percent of NSCLC patients carry oncogenic KRAS mutation[2]. A potent drug targeted against KRAS mutation has not yet been developed and the objective response rate with the current standard of care is just three percent. An earlier report had suggested siRNA therapy renders the undruggable KRAS mutant cells to become susceptible to Tyrosine Kinase Inhibitors (TKI)[3]. Short interfering RNA (siRNA) is a well-known approach for effecting gene therapy to provide subsequent sensitization towards complementary therapeutic agents. However, stable delivery of siRNA is a significant challenge due to its high degradation rate in the presence of serum proteins and enzymes. To overcome this challenge, several nanoparticle based carrier systems have been attempted and those include retroviral vectors, liposomes, polymeric, and metallic nanoparticles[3–7]. In these reported studies the physicochemical and surface properties of the particle were modified for delivering the siRNA to cytoplasm of the infected cells. Unfortunately, these nanoparticles suffer from serious limitations such as stability issues during synthesis, premature release in serum, inefficient endosomal escape, and interferon response[4,8,9]. Importantly, oncogene knockdown alone has less impact on the cancer cell apoptosis since the cells tend to adopt another effector pathway for survival[3,10–12]. Therefore, a need for complementary drug for initiating the apoptosis post knockdown is needed. Indeed, drugging cells separately and exogenously post oncogene knockdown has been reported earlier[9–11]. A combined delivery system wherein, co-delivery of a drug along with siRNA to impede growth and survival of the cell has also been attempted[13]. The relevance of the combined delivery is to ensure the complementary drug enters the same cells that are affected by siRNA at a predetermined appropriate proportion and time for causing cellular apoptosis. However, incorporation of siRNA (with minimal degradation) with a drug and a biomarker-targeting antibody into a single platform is synthetically challenging. Thus, stable and targeted delivery with concomitant cytotoxic action to cancer cells continues to be at early exploratory stages.

[1]Department of Radiology, Medical Sciences Building, University of Missouri, Columbia, MO 65212, USA. [2]Department of Bioengineering, University of Missouri, Columbia, MO 65212, USA. [3]Department of Pathology, Medical Sciences Building, University of Missouri, Columbia, MO 65212, USA. [4]Notre Dame Integrated Imaging Facility, University of Notre Dame, Notre Dame, IN 46556, USA. [5]Department of Medical Pharmacology and Physiology, University of Missouri, Columbia, MO 65212, USA. [6]Institute of Clinical and Translational Science, University of Missouri, Columbia, MO 65212, USA. [7]International Center for Nano/Micro Systems and Nanotechnology, University of Missouri, Columbia, MO 65212, USA. Correspondence and requests for materials should be addressed to R.K. (email: kannanr@health.missouri.edu)SCIENTIFIC REPORTS | 6:30245 | DOI: 10.1038/srep30245    1

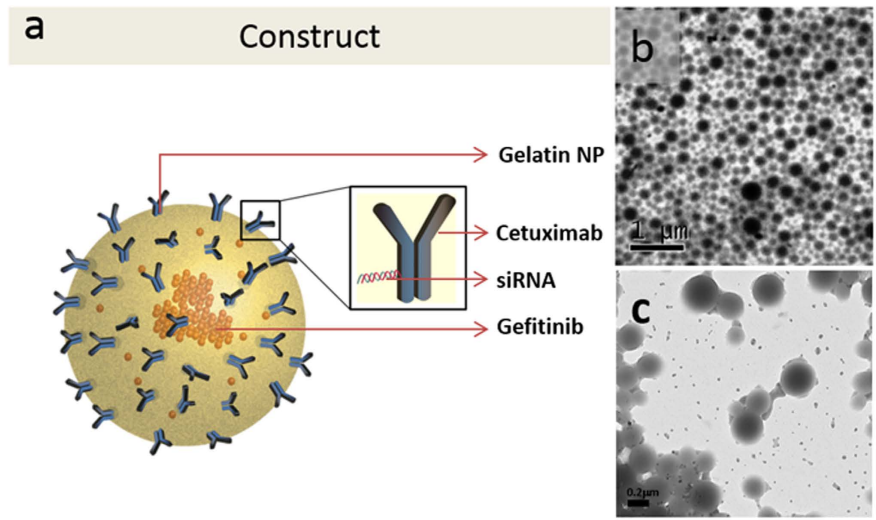

**Figure 1.** (**a**) Schematic representation of tri-block nanoparticle (TBN) consisting of gelatin nanoparticle encapsulated with gefitinib and surface functionalized with cetuximab conjugated siRNA; (**b**) TEM image of Gelatin Nanoparticles; and (**c**) TEM image of TBN.

Significant efforts have been made to understand the downstream effect of oncogene knockdown mediated via siRNA till date[14]. Cancer cells have several parallel working pathways, with one primary effector pathway coupled to several parallel effector pathways[15]. The parallel pathways remain dormant until the working pathway is disrupted. Change in the protein expression levels upon knock down of oncogene present in the primary pathway results in change of downstream protein and gene expression levels regulated by complex cellular mechanism. This mode of intra-cellular functioning adaptation evolves to drug resistance within cancer cells that are previously responding to therapy[16]. On the other hand, KRAS mutant adenocarcinoma of NSCLC have been undruggable till date[17]. While mutations occur at variation position of KRAS, oncogenic effect at codon 12 (Glycine-12 to Cysteine, G12C) of KRAS is the most commonly occurring mutation and yet to receive a dedicated drug[18]. Although, in recent times, few attempts have been made for targeting G12C mutation through a small molecule inhibitor, RNAi therapy is emerging as a promising tool that could be applied across all types of mutations supplemented with currently approved drugs[19,20]. For example, knocking down a specific gene of undruggable cancer, such as KRAS mutant adenocarcinoma of NSCLC, can activate a parallel dormant effector pathway that may be sensitive to a TKI[3].

In this work, we report the synthesis and utilization of gelatin nanoparticle (Gel NP) as a carrier system encapsulated with gefitinib (Gel$_{GEF}$NP). The Gel NP is surface functionalized with cetuximab (Ab), a EGFR targeting antibody, conjugated to KRAS G12C specific siRNA (Ab-siRNA conjugate) (Fig. 1). KRAS G12C specific siRNA is chemically conjugated to cetuximab by thio-ether linkage. Cetuximab antibody target EGFR receptor on the cells and also protects siRNA from external degradation. We believe the 14 kDa siRNA is camouflaged by 146 kDa cetuximab antibody thus remaining un-exposed to external environment. The knockdown of oncogene upon treatment with TBN was confirmed by studying downstream signaling events studying protein expression levels, and the effect of sensitization towards drug molecule was determined by *in vitro* cytotoxicity assays. A possible mechanism governed by GAB1 mediated downstream signaling as an adopted pathway post oncogene knockdown has been shown. Change in expression levels of two cytoplasmic genes confirming oncogene knockdown and the mechanism of sensitization towards a TKI due to the adopted effector pathway is reported in this study.

## Results

**Synthesis of Tri-block Nanoparticle (TBN).** Gefitinib encapsulated Gel NPs (Gel$_{GEF}$NP) were prepared using two-step desolvation method[21,22]. Gel$_{GEF}$NP was washed several times to ensure no diffusion or desorption mediated release of gefitinib occurs and that any release occurring from the nanoparticles is attributed only to degradation assisted release mechanism. For conjugating cetuximab monoclonal antibody on the surface of Gel$_{GEF}$NP, carboxyl groups present on the surface were converted to reactive NHS ester followed by addition of antibody. The amine groups present in antibody reacted to the modified carboxyl group enabling the conjugation of the antibody to Gel$_{GEF}$NP. Subsequently, amine-thiol linker sulfo-SMCC, was used to convert amino groups on the antibody to thiol reactive maleimide group. KRAS specific siRNA with 5′ sense strand modified with disulfide end group was reacted to maleimide present on the antibody for the preparation of antibody-siRNA conjugate, thus forming the tri-block nanoparticle (TBN) (see supplementary information Figures S-1 and S-2).

**Characterization of Tri-block Nanoparticle.** Physical properties of Gel NPs prepared via desolvation process conformed to earlier reports[21,22]. Gel NPs dispersed in DI water indicated an average hydrodynamic diameter of $225 \pm 20$ nm with a polydispersity index (PDI) less than or equal to 0.1. Zeta potential ($\zeta$) measurements of Gel NPs suspended in DI water showed a positive value of $+19$ mV ($\pm 2$). As expected, the size and $\zeta$ did not change with gefitinib encapsulation indicating surface properties of Gel NPs remain unchanged with encapsulation (see supplementary information Figure S-3). TEM image analysis displayed a core size similar to the measured hydrodynamic diameter (~200 nm) (Fig. 1).

To determine the encapsulation efficiency of gefitinib within Gel NPs and percent of drug loading, UV-Vis absorption spectroscopy was used. In this experiment, $Gel_{GEF}NP$ at known concentration dispersed in DI water and subjected to protease degradation. The translucent solution turned transparent after 1 hour indicating digestion of gelatin matrix. To ensure that the particles are completely degraded, the solution was centrifuged at 20,000 g for 20 minutes and absence of precipitate indicated complete degradation of the matrix. The solution was then passed through 0.2 $\mu$m filter and the filtrate was analyzed for characteristic 331 nm absorption peak of gefitinib[23]. Gefitinib calibration curve was used as reference. Analysis revealed 40% encapsulation efficiency with 5 $\mu$g of gefitinib per mg of Gel NPs (0.5% loading, post all release including burst release and diffusion/desorption mediated release).

To determine the amount of monoclonal antibody, cetuximab, conjugated to the surface of Gel NPs or $Gel_{GEF}NPs$, Bradford protein estimation assay was used. The assay showed that 45 ($\pm 5$) $\mu$g of cetuximab was conjugated per mg of Gel NPs. The percentage of antibody conjugation, in other words, conjugation efficiency changed proportionally with the relative initial amount of nanoparticles (Gel NPs or $Gel_{GEF}NPs$) for all 5 experiments carried out for reproducibility (amount of cetuximab was kept constant at 1200 $\mu$g). However, amount of antibody conjugated per mg of nanoparticles remained constant at 45 ($\pm 5$) $\mu$g per mg of Gel NPs or $Gel_{GEF}NPs$. The hydrodynamic size of the antibody-conjugated nanoparticles (Ab-Gel NPs or Ab-$Gel_{GEF}NPs$) did not show any significant change and was found to be 225 ($\pm 15$) nm. TEM images however appeared relatively well resolved in the case of Ab-Gel NPs/$Gel_{GEF}NPs$ compared to that of Gel NPs or $Gel_{GEF}NPs$. (Fig. 1 and Figure S-2) This could be due to the increased surface density of the nanoconstruct leading to a relatively higher contrast for the Ab-$Gel_{GEF}NPs$ compared to that of Gel NPs or $Gel_{GEF}NPs$. The $\zeta$ of Gel NPs and Ab-Gel NPs, however, showed a drastic change post surface modification; for example, the $\zeta$ changed from $+18$ mV to $-12$ mV confirming alteration on the surface (Figure S-3).

To estimate the concentration of siRNA loaded on the Gel NP or $Gel_{GEF}NP$, florescence spectroscopy was used. In this experiment, Cy5 labelled siRNA was used as a surrogate to estimate the loading of siRNA. After conjugation reaction, the supernatant after centrifugation (20,000 g, 20 mins) was analyzed for characteristic 660 nm emission peak of cy5 using florescence spectroscopy. Using the manufacturer's extinction coefficient (367,569 L. $mol^{-1}.cm^{-1}$), the fluorescence signal obtained was compared with reference curve of $siRNA_{cy5}$ and the conjugation of siRNA on the nanoparticle was determined to be ~98%. Visual observation of characteristic blue color of Cy5 present on the nanoparticle pellet and absence of the color in the supernatant also confirmed the high siRNA conjugation qualitatively (see supplementary information Figure S-4). Absolute number of siRNA, antibodies and TKIs per Gel NP is listed in Table S-1.

**Stability of Tri-block Nanoparticle.** To estimate the serum stability of unmodified siRNA and siRNA conjugated to the nanoparticle, we utilized native polyacrylamide gel electrophoresis (PAGE) technique as reported previously[24]. The nanoparticles with appropriate controls including naked siRNA and Ab-Gel NP were added to 10% serum solution at 37 °C. At predetermined time intervals (0, 0.5, 1, 2, 4, 8 and 24 hr), 100 $\mu$l of each sample was removed and stored at $-50$ °C. The solutions were thawed to room temperature and subjected to native PAGE; the presence of siRNA was detected using nucleic acid staining GelRed dye (Fig. 2a). The use of thio-ether bond for conjugating siRNA to the antibody present on the gelatin nanoparticles renders it uncleavable from the nanoconstruct[25]. Any gelatin degradation agent such as protease or antibody disruption would have adverse effect on siRNA quality and indicate false-negative results. Therefore, the stability study of the siRNA was based on the direct GelRed nucleic acid staining of the siRNA present on the nanoparticle as reported earlier[26]. Naked siRNA and Ab-Gel NPs/$Gel_{GEF}NPs$ were used as controls to evaluate the 24 hr stability of siRNA. Unbound or electrostatically bound siRNA to nanoparticles had degradation profile similar to naked siRNA. However, siRNA conjugated to nanoparticles were highly stable with minimal degradation for 24 hours. In addition, 30 days stability analysis of the nanoparticle stored at $-50$ °C showed minimal degradation (Fig. 2b). *In vitro* stability studies of TBN as analyzed by hydrodynamic size in biological media containing 10% serum showed no significant variations in the diameter of the nanoparticles until 24 hr time points (Fig. 2c).

**Cellular Internalization of Tri-block Nanoparticle.** To understand the cellular internalization capability, we used Gel NP encapsulated with fluorescein (fl) dye ($Gel_{fl}NP$) and siRNA labeled with cy5 dye ($siRNA_{cy5}$) for detection using two different fluorescent signals. The internalization study was performed in H23 cancer cell line and analyzed using fluorescence microscopy (Fig. 3). Indeed, co-localization of $siRNA_{cy5}$ and $Gel_{fl}NP$ was confirmed by two fluorescence signals emanating from cells. To estimate the relative amount of siRNA internalization within cells, we performed flow cytometry analysis. In this experiment, we incubated H23 cells independently with (i) $siRNA_{cy5}$ along with transfecting agent (TA-$siRNA_{cy5}$); (ii) Gel NP-Ab-$siRNA_{cy5}$; or (iii) Ab-$siRNA_{cy5}$ for 4 hours (Fig. 4). After 4 hours, the cells were repeatedly washed to remove surface adhered molecules and were trypsinized. Subsequently, the cells were analyzed using flow cytometry and the results showed internalization in the following order: Gel NP-Ab-$siRNA_{cy5}$ > TA-$siRNA_{cy5}$ > Ab-$siRNA_{cy5}$. As expected, the amount of siRNA internalized through nanoparticle was at least two fold higher than transfected siRNA (Fig. 4). Results showed the superiority of TBN as an effective siRNA delivery system.

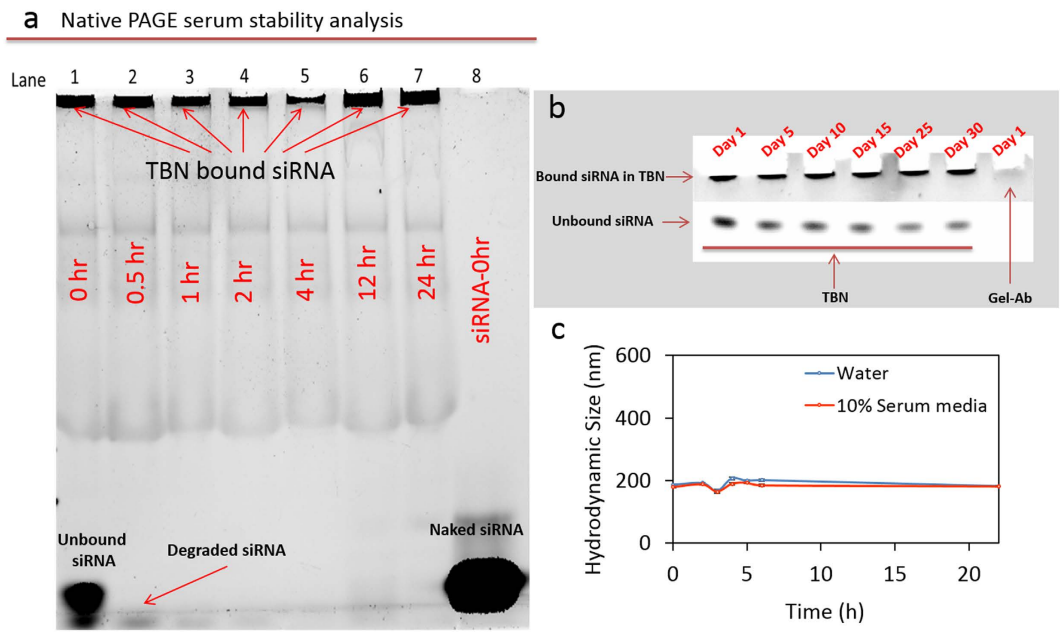

**Figure 2. Stability of siRNA and TBN in serum.** (**a**) Serum stability with time was analyzed using native PAGE, the results reveal that siRNA present on TBN are highly stable. Lane description – 1: 0 hr; 2: 0.5 hr; 3: 1 hr; 4: 2 hr; 5: 4 hr; 6: 12 hr; 7: 24 hr; and 8: naked siRNA at 0 hr in DI water. The lower band in lane 1 corresponds to free siRNA and siRNA cleaved from Tri-block nanoparticle (TBN) by serum solution. The band disappears within 30 min. indicating the short stability of free siRNA in serum; (**b**) Long term siRNA stability analysis of nanoparticle stored at −50 °C in presence of cryoprotectant; and (**c**) *In vitro* stability of TBN in biological media containing 10% serum at various time points analyzed by measuring hydrodynamic diameter of nanoconstruct.

**Knockdown Mediated Protein Regulation.** It is important to ensure that internalized nanoparticle is present in the cytoplasm and knocks down the KRAS gene as intended and regulates the appropriate protein levels. In order to understand downstream protein level regulation before and after oncogene knockdown, we used H23 cell line harboring KRAS mutation at G12C. In this experiment, we studied the effect of TBN on RAS/RAF/MEK/ERK cascade, a complex signal transduction pathway that has been extensively studied and mapped[27]. Untreated H23 cells exhibits phosphorylated downstream proteins in two effector pathways. In first pathway, RAS activates downstream effector enzymes enabling cell proliferation and survival through phosphorylation of RAF. This phosphorylation in-turn activates mitogen activated protein kinase (MEK/MAPK) that is responsible for activation of ERK. In second effector pathway, RAS has been found to activate PI3K effector pathway leading to phosphorylated AKT[15]. PI3K/AKT signaling network that runs parallel to RAS/MAPK pathway is known to have several points of interaction with each other influencing inter-signal transduction[15]. Therefore, disruption of RAS with oncogene knockdown should result in downregulation of primary downstream protein pMEK as well as parallel pathway protein pAKT. Indeed, the results indicate a direct intrinsic relation between the two pathways primarily governed by RAS (Fig. 5). Similar results were obtained in previous studies; for example, oncogene knockdown using retroviral shRNA in H23 resulted in significant reduction of downstream-activated proteins including pMEK, pERK and pAKT[3]. In addition to the above pathways, it is also interesting to know the status of EGFR in these KRAS mutation cell lines. Previous studies have already established that mutations in KRAS and EGFR are mutually exclusive[28]. Therefore, all EGFR downstream proteins will largely be dormant. However, a recent study showed that post knockdown of KRAS oncogene showed the presence of pEGFR within three days suggesting activation of EGFR to compensate for the loss of KRAS oncogenic activity[3]. Similarly, inhibition of activated ERK sensitized TKI resistant cells towards gefitinib[29]. Another report suggested a similar outcome wherein, blocking PI3K and MAPK signaling pathways led to gefitinib sensitization for TKI resistant NSCLC[30]. Even though progress has been made to unravel the underlying mechanism that enables sensitization of KRAS cells towards TKIs, it remains largely inconclusive. In this regard, Wu and coworkers recently reported the dissociation of "SHP2-GAB1" proteins within KRAS mutant cells, when treated with erlotinib[31]. The study further showed inhibition of GAB1 tyrosine phosphorylation in HCC827 cells, a TKI responsive cell line, when treated with erlotinib. Based on these studies, we focused to investigate a possible GAB1 mediated survival pathway in the case of treatment of TBN with H23 cell line. Our results showed that GAB1 and pGAB1 at Tyr 627 was found to be upregulated upon treatment with TBN while no substantial difference was found for pGAB1-307 (supplementary information Figure S-5). Also, SHP2, pSHP2 (Y542) and pSHP2 (Y580) remained unchanged for Gel NP (without gefitinib) but was found to be downregulated in the presence of gefitinib for the cells transfected with siRNA or treated with TBN (Fig. 6). Considering the downstream protein regulation was significantly altered through siRNA mediated gene therapy, we extended our investigation in understanding whether TBN would render the cells susceptible to TKI.

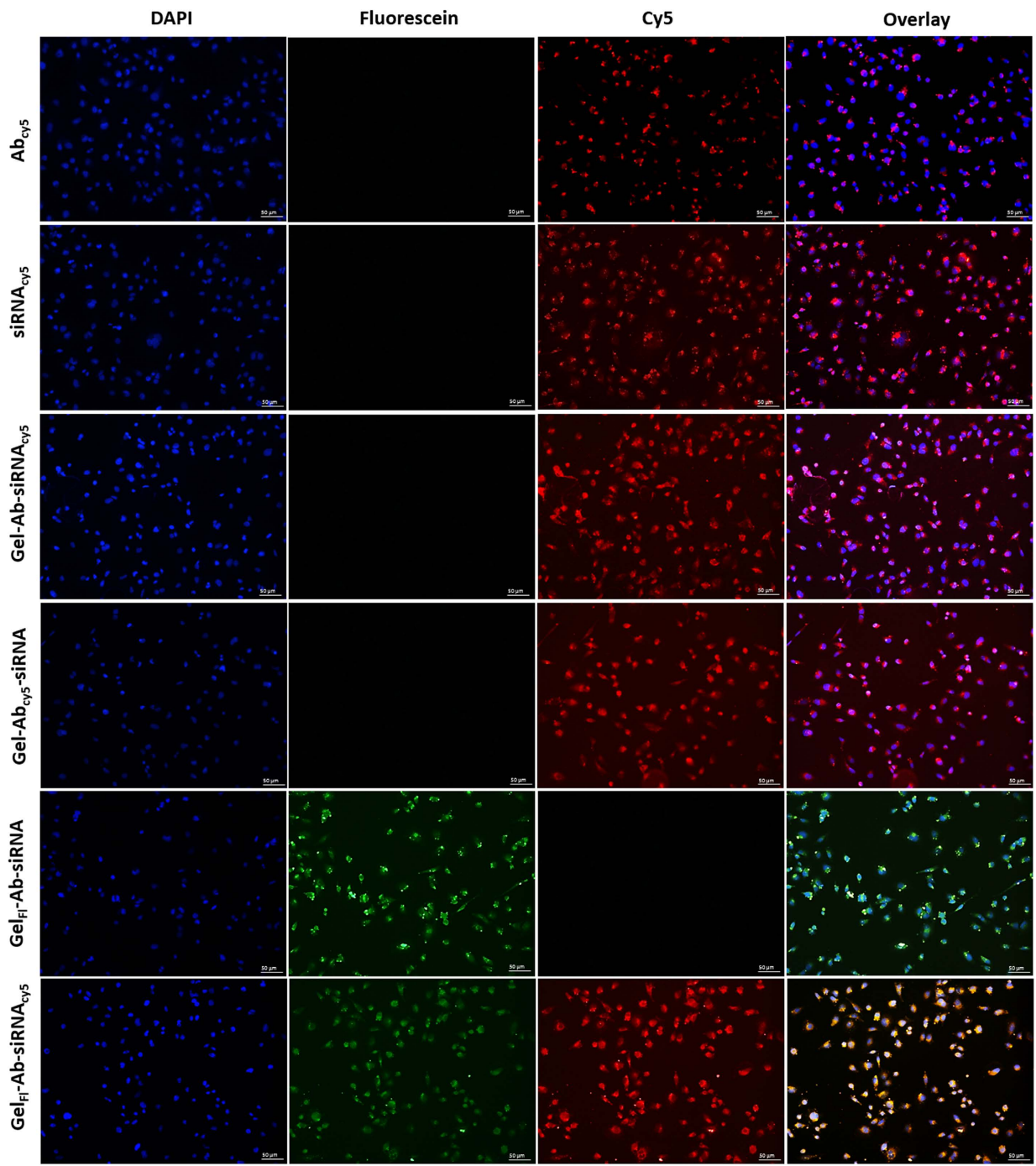

**Figure 3. Cellular localization of TBN in H23 cells.** Florescence microscopy of H23 cells incubated with TBN (Gel$_{Fl}$Ab-siRNA$_{Cy5}$) and its analogues Gel$_{Fl}$Ab-siRNA, Gel-Ab$_{Cy5}$-siRNA, Gel-Ab-siRNA$_{Cy5}$, siRNA$_{Cy5}$ and Ab$_{Cy5}$. DAPI was used as a nuclear staining marker. Fluorescein was used for encapsulation in Gelatin Nanoparticles. Cy5 was used for labelling antibody and siRNA. The figure shows co-localization of cy5 labelled siRNA and fluorescein encapsulated gelatin nanoparticles in H23 cells. Appropriate controls with and without fluorescein or cy5 were used. All images were recorded at 20X magnification.

**Efficacy of Tri-block Nanoparticle.** Considering downregulation of downstream-phosphorylated proteins, our next step was to investigate the cell viability of H23 cells (MTT assay) after knocking down RAS pathway. For comparison of various individual components of the nanoparticle, toxicity data was normalized with gefitinib concentration and the corresponding concentration of appropriate controls are shown in the supplementary information Figure S-6. IC$_{50}$ value of gefitinib for H23 cell lines was determined to be 50 μM. Interestingly, transfected siRNA (4 μM, dose corresponding to 50 μM gefitinib) did not cause any cytotoxicity and the viability of the cells remained 100% suggesting oncogenic disruption alters effector pathway but does not lead to apoptosis. On the other hand, when we used TBN the IC$_{50}$ value drastically reduced by 20 fold to 2.5 μM, and complete loss of cell viability was found for nanoparticle containing 5 μM of gefitinib (Fig. 7). A similar cytotoxicity trend was found when H23 cells were transfected with siRNA and gefitinib was added exogenously. The controls used for cytotoxicity assay include gefitinib, transfected siRNA with and without gefitinib (siRNA (TA)+gef, siRNA (TA)), cetuximab (Ab), cetuximab-siRNA (Ab-siRNA) with and without gefitinib, gelatin nanoparticles (Gel NP), gefitinib encapsulated gelatin nanoparticles (Gel$_{GEF}$NP), and mock siRNA with gefitinib (see supplementary information Figures S-7–S-9). No significant changes in toxicity relative to nanoparticle were found in the case of controls.

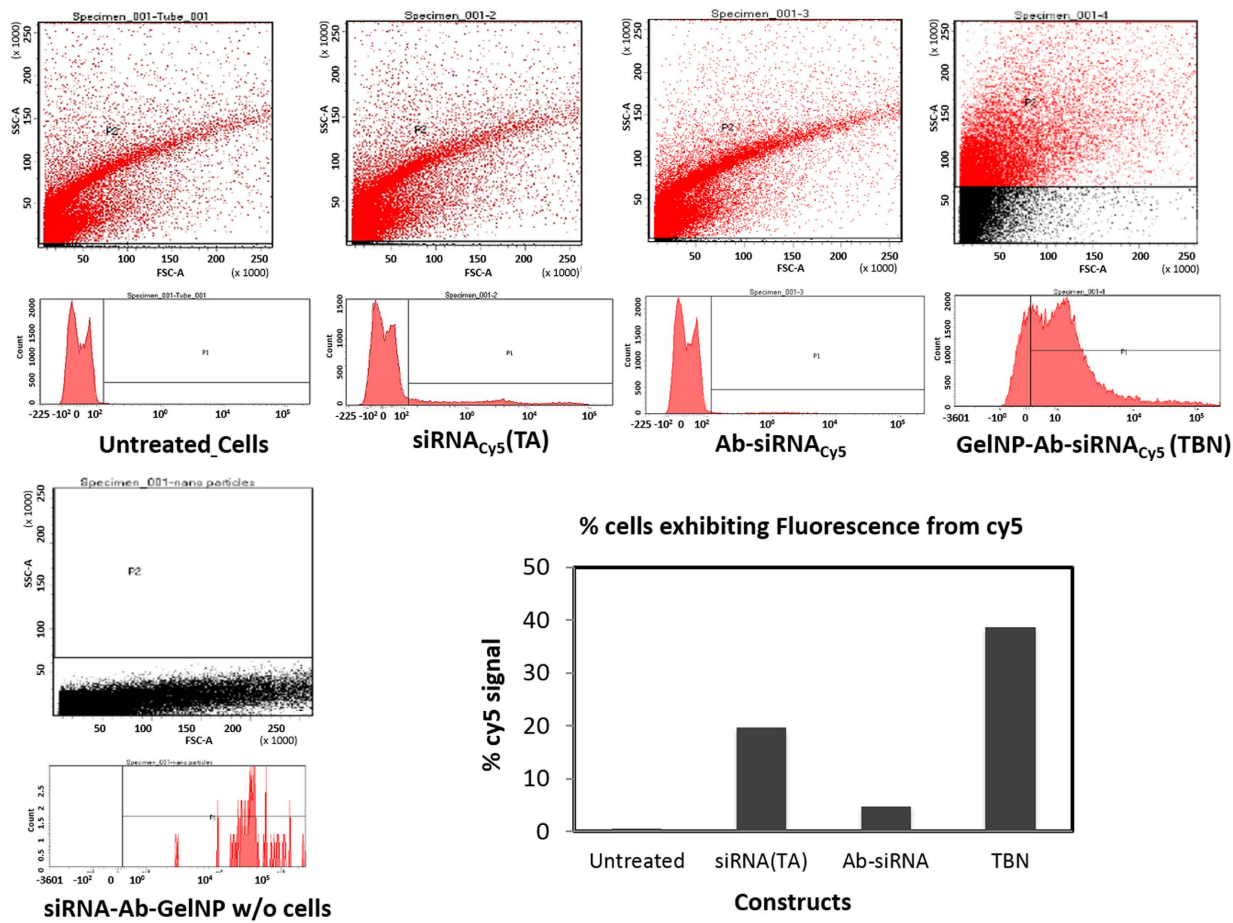

**Figure 4. Flow cytometry analysis.** Flow cytometry analysis reveals GelNP-Ab-siRNA$_{cy5}$ internalizes almost 2 folds compared to transfected siRNA. Cetuximab-siRNA (Ab-siRNA), on the other hand, has minimal internalization relative to other compounds.

**Gene Regulation.** To understand the effect of siRNA knockdown and protein down-regulation on the effector pathway adopted for survival of cells, we performed quantitative real time PCR by examining dual specificity phosphatase 6 (DUSP6) and CD73 gene expression levels. Earlier reports have indicated that DUSP6 expression is regulated by ERK signaling in RAS/RAF/MEK/ERK cascade[32]. Also, DUSP6 is an important feedback loop as it exhibits antitumor profile through negative feedback regulation[33]. An effect on the signaling cascade, therefore, must have an effect on DUSP6 levels. Transfected siRNA and TBN showed downregulation of DUSP6 upon knockdown. The RF values of DUSP6 post knockdown determined using real time qPCR for nanoparticle correlated well with that of transfected siRNA and was determined to be 0.17 and 0.1 for TBN and siRNA respectively. In contrast, Ab-Gel$_{GEF}$NP devoid of siRNA, showed the RF value of 0.73 indicating minimal or no effect on the gene regulation. The results suggest post oncogene knockdown, loss of activity in the MEK/ERK cascade has a direct impact on the gene regulation of DUSP6 (Fig. 8a). NT5E or CD73 is a gene that has shown relevance as a predictive biomarker for overall survival and progression free survival in human patients harboring KRAS wild type or KRAS mutation. In another study, NT5E transfected breast cancer T-47D cells resulted in increased levels of cell migration and invasion[34]. Also, significant increase in protein expression levels of EGFR was observed post transfection. It was also found that CD73 expression is directly proportional to EGFR expression and suppression of EGFR resulted in decreased CD73 levels. These observations led us to investigate the CD73 gene level expression for understanding the mechanism post loss of activity on RAS/RAF/MEK/ERK cascade after siRNA mediated knockdown. Interestingly, we observed a significant increase in the CD73 expression levels for cells treated with TBN and no change for the cells transfected with siRNA or treated with Ab-Gel$_{GEF}$NP devoid of siRNA. The CD73 RF value for cells treated with TBN was determined to be 3.64 compared to 1.07 and 0.95 for siRNA and Ab-Gel$_{GEF}$NP with no siRNA respectively (Fig. 8b). The increase in CD73 expression levels could be attributed to presence of both siRNA and cetuximab in TBN.

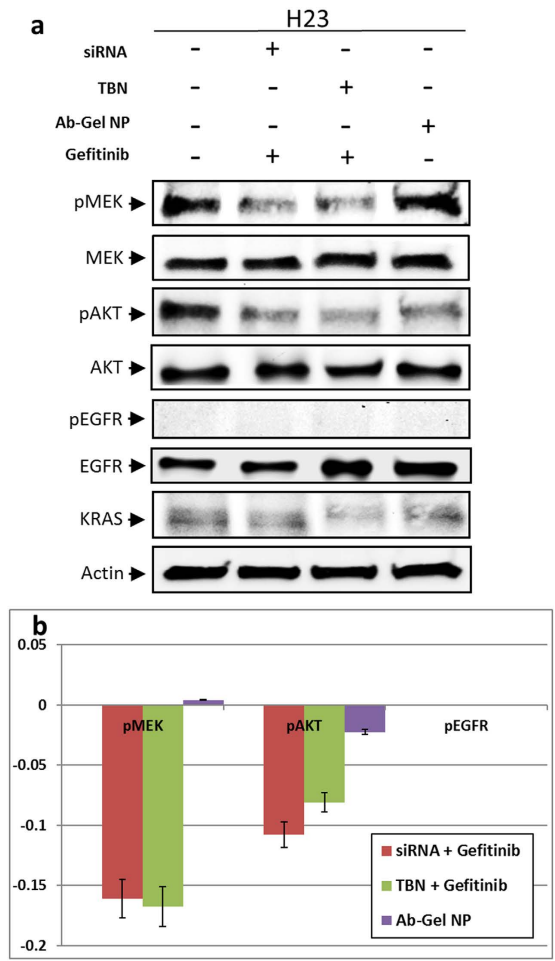

**Figure 5.** (**a**) Western Blot analysis showing the effect of KRAS mutant knockdown by transfection agent or through TBN and gefitinib on the phosphorylation of MEK, AKT, EGFR in the KRAS mutation positive NSCLC cell line; (**b**) A graphical representation using densitometry image analysis, showing the protein expression levels after KRAS knockdown. The data were obtained from three independent experiments.

In summary, based on the collective data obtained we were able to predict the mechanism of release of TBN from endosomes to cytosol (Figure S-10) and the concomitant effect of gene and protein regulation. As an additional control, we studied the effect of siRNA and TBN in A549 NSCLC cells that has KRAS G12S mutation (Figure S-11). It is worth to note the siRNA that we used in our present study is targeted for G12C mutation. Based on the cell viability data, it is evident that siRNA and TBN is very specific for KRAS G12C mutation and have minimal or no effect of G12S mutation. The results further confirm the specificity of the conjugate toward particular mutation.

***In vivo* dose safety study.** We performed a preliminary evaluation of *in vivo* safety of the TBN in normal mice. The safety study was conducted in five normal mice by repeated intravenous (IV) injection of the nanoconstruct (80 mg/Kg body weight) for three consecutive days, followed by euthanasia. Subsequently, major organs were collected and histopathology was performed. The data showed that animals did not show any abnormal behavior, and histology analysis indicated no signs of toxicity showing tolerance of injected dose in animals (Figure S-12).

## Discussion

We synthesized a targeted TBN as an effective siRNA delivery system to sensitize the KRAS mutant tumor cells to gefitinib. The study demonstrated that camouflaging siRNA between gelatin and antibody molecules in TBN increased the stability of the siRNA. Indeed, serum stability and cytoplasmic delivery of siRNA present within TBN are attributed to synergistic effect of both Gel NP and cetuximab antibody. The release of encapsulated gefitinib from TBN primarily governed by degradation of gelatin matrix of the nanoparticle; however, exposure or release of siRNA to form RNA-induced silencing complex (RISC) can occur in two ways, either through the surface degradation of Gel NP that subsequently releases Ab-siRNA conjugate forming RISC or via direct complexation of RISC with siRNA present on the nanoparticle (Figure S-10). However, determining the dominant mechanism that is responsible for complexing with RISC is difficult to detect or predict.

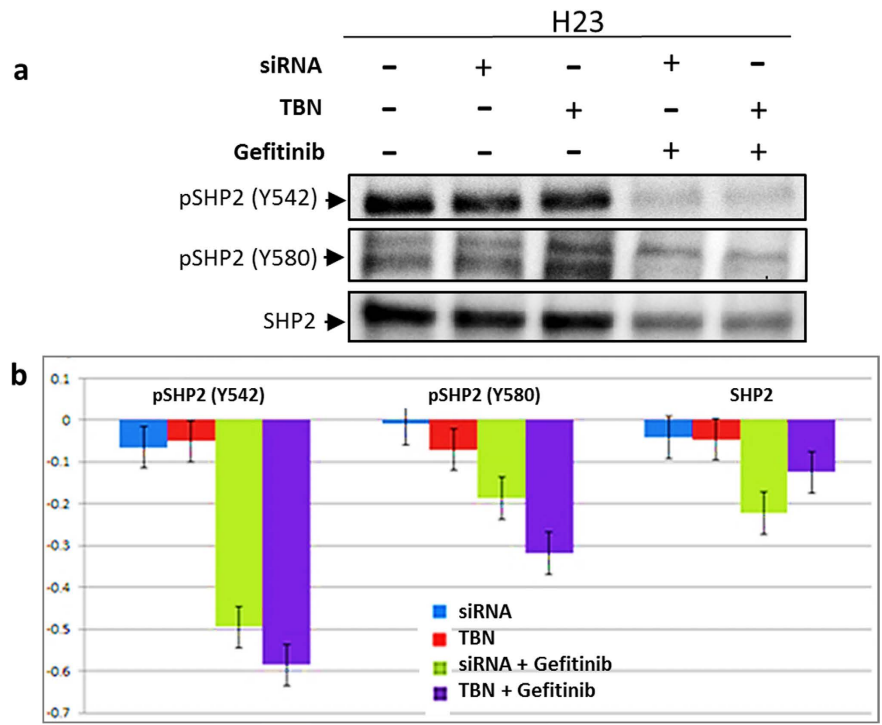

**Figure 6. Gefitinib induced protein regulation.** (**a**) Treatment on KRAS mutated H23 cells by transfected siRNA and siRNA delivered using TBN did not show any effect on SHP2 and p-SHP2. However, in the presence of gefitinib, pSHP2 (Y542), pSHP2 (Y580) and SHP2 protein were downregulated indicating impaired SHP2 function – a probable cause of H23 cell apoptosis after oncogene knockdown; (**b**) Densitometry image analysis using BioRad Image lab V.3 was carried out to quantify the proteins.

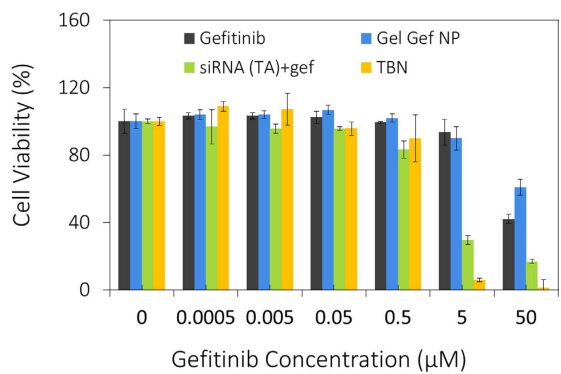

**Figure 7. *In vitro* cytotoxicity.** Cytotoxicity data indicates nanoparticle treatment knocks down oncogene and sensitizes KRAS mutant H23 cell line towards gefitinib. The cell viability of TBN at 5 μM gefitinib concentration is 5% compared to 25% viability for cells transfected with siRNA followed by gefitinib treatment.

As reported in the literature, siRNA mediated mutant KRAS oncogene knockdown results in down regulation of phosphorylated proteins present downstream of RAS pathway[3]. In our study, we observed a similar trend, wherein pMEK present in RAS pathway was downregulated by mutatnt KRAS oncogene knockdown. In comparison with untreated cells, the relative down regulation of pMEK was found to be at least 2 times lower when we used TBN. As expected, the down regulation of the downstream proteins mediated by TBN showed similar response as that of transfected siRNA. Gel NP without siRNA had no effect on pMEK, thereby proving that knocking down KRAS oncogene effectively decreases RAS functioning pathway. Also, RAS/MEK/ERK pathway is known to be closely associated to PI3K/AKT pathway. With 802 interactive proteins involved in PI3K signaling and over 2000 proteins in the case of MAPK pathway, several cross talk points exist between PI3K/AKT and RAS/MAPK pathway[15]. Among these proteins, Grb2 associated binder-1 (GAB1) has been identified as an important adaptor protein playing central role in various cell responses[35,36]. Specifically, GAB1 is found to be functionally active docking protein for several downstream signaling pathways including EGFR. PI3K and GAB1 share intrinsic association via tyrosine domain phosphorylation of GAB1 on SHP2 binding motifs. SHP2 negatively regulates PI3K activation through dephosphorylation of GAB1 phosphotyrosinases facilitating the activation of RAS for certain cases[37]. In others, activated ERK results in phosphorylation of GAB1 (Grb2 associated binder 1) on serine and threonine residues adjacent to p85 PI3K binding sites and the nature of signaling dictates regulation levels of GAB-p85 PI3K complexes to control PI3K activity[15]. In the case of H23 cells, we found that pAKT reduced with downregulation of pMEK and pERK. The result suggests that the feedback loop of MAPK with PI3K is possibly governed by GAB1 phosphorylation on serine and threonine residues and not via SHP2-PI3K binding prior to RAS oncogene knockdown. The silencing of activated ERK signaling by RAS oncogene knockdown disrupts the formation of GAB-P85 PI3K complex by negatively affecting serine and threonine phosphorylation levels of GAB1. This negative feedback results in arresting PI3K pathway and downregulating AKT activity in the case of H23 cells. A similar result was reported previously, that is ERK positively regulate GAB1 p85 interaction; subsequently inhibiting ERK and result in decreased AKT activity[38,39]. It is known that ERK regulation of GAB P85 PI3K complexes is mediated by Hepatocyte Growth Factor (HGF)[35]. The fact that HGF is highly expressed in the case of H23 cells strengthens our claims that ERK induces positive regulation of GAB1 for associating with P85-PI3K via HGF mediation under normal conditions, disruption of which deactivates PI3K/AKT pathway[40,41].

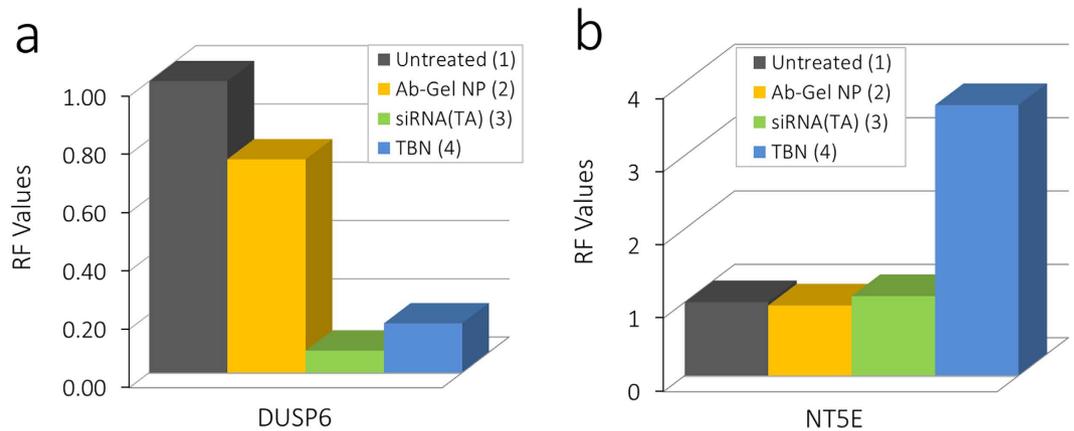

**Figure 8. RT-PCR analysis.** (**a**) RF values determined by quantitative real time PCR indicates down regulation of DUSP6 gene expression for the H23 cells treated with siRNA and TBN; (**b**) RF values determined by quantitative real time PCR indicates up regulation of NT5E gene expression for the H23 cells treated with TBN. Cells treated with siRNA or cetuximab functionalized gelatin nanoparticles showed minimal or no change in expression levels compared to the untreated cells.

Interestingly, there was no effect on cell viability with 100% cell survival with disruption of RAS/MEK/ERK pathway post RAS oncogene knockdown (for both transfected siRNA and Gel NP without gefitinib). Addition of gefitinib to siRNA treated cells, however, showed considerable effect on the viability of the cells, more so for the TBN compared to transfected siRNA. The results corroborate well with an earlier report wherein, Hoeben *et al*. showed inhibition of GAB1 activity in head and neck squamous cell carcinoma reduced signaling of EGFR resulting in sensitization of the cells towards gefitinib[42]. Several other studies have shown intricate relationship existing between tyrosine kinase inhibitors and disruption of MAPK/PI3K pathways[30,43,44]. However, the question that still needs to be addressed is the functional mechanism of the H23 cells for survival with knocked-down RAS oncogene, and in effect, with arrested MAPK and PI3K pathways. As mentioned earlier, tyrosine phosphorylation of GAB1 can effectively associate with EGFR. With inability of GAB1 to form complexes with p85 PI3K due to loss in serine phosphorylation, GAB1 can effectively regulate EGFR signaling through several positive feedback loops. One possibility is that association of GAB1 with EGFR by recruiting SHP2 and cascading EGFR downstream signaling. In the case of downstream EGFR, SHP2 is activated by complexing with phosphorylated GAB1 that binds with EGFR through Grb2. Tyr 627 domain is one of the predominant domains of pGAB1 wherein the complexing occurs[45]. Indeed, western blot results indicated an increase in the pGAB1 (Tyr 627) after oncogene knockdown by TBN supporting our hypothesis that knockdown of KRAS oncogene arrests MEK/ERK pathway, triggering a feedback loop wherein GAB1 dephosphorylation on serine and/or threonine residues occur, resulting in abrogation of AKT activity. The stimulation cascades to phosphorylation at tyrosine domains of GAB1 that induces association of GAB1 to EGFR through SHP2 recruitment. It is thus possible that the adopted effector pathway of survival post oncogene knockdown is governed through EGFR downstream signaling *via* disrupted MAPK pathway. It is noteworthy to mention that SHP2 regulation has intrinsic relation with tyrosine kinase inhibitors. It has been reported that impaired SHP2 functioning or altered localization of SHP2 causes sensitivity to gefitinib[31,46,47]. In our case, the results suggest that alteration of SHP2 localization caused by recruitment through GAB1 causes disruption of EGF downstream signaling. Addition of gefitinib impairs the functioning of SHP2 and disables complex formation with GAB1, thereby abrogating the already disrupted MAPK pathway, leading to apoptosis of the cells (Fig. 9).

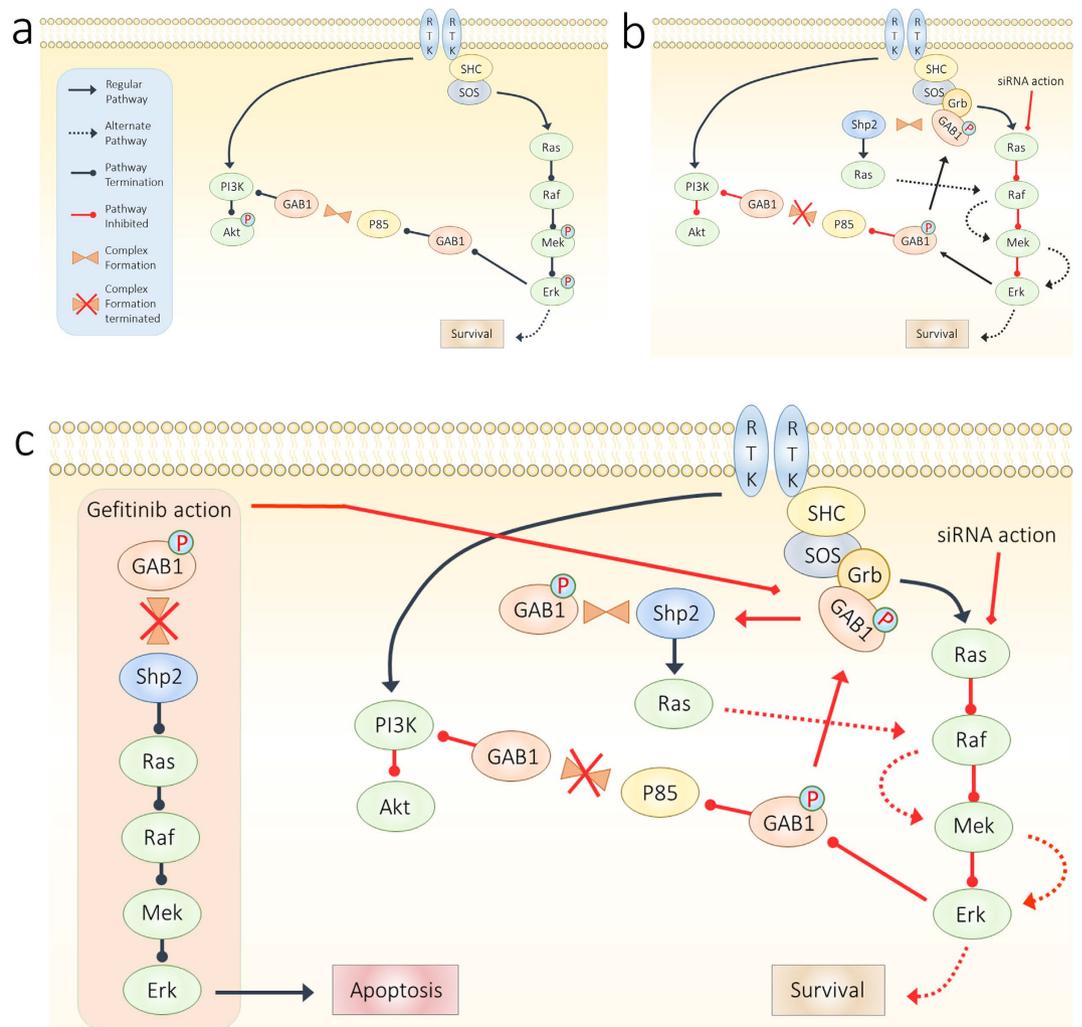

**Figure 9. Effect of siRNA and Gefitinib on downstream protein regulation mechanism.** (**a**) Effector pathway in the presence of oncogenic RAS mutation in H23 cells; (**b**) disruption of RAS downstream through siRNA mediated oncogene knockdown affects regular functioning of RAF/MEK/ERK pathway leading disrupted functioning through MAPK pathway caused by alerted localization of impaired SHP2; and (**c**) Addition of gefitinib impedes the recruitment of SHP2 by GAB1 leading to the abrogation of the already disrupted MAPK pathway and eventual apoptosis of the affected cells.

In view of the pathway being governed by EGFR signaling for cells with knocked down mutant RAS oncogene, it was of interest to understand the behavior of gene regulation considering the nanoparticle contains cetuximab, an EGFR inhibitor. Transfected siRNA, however, could serve as a positive control in terms of oncogene knockdown but a negative control for EGFR blocking. Conversely, nanoparticle without siRNA could serve as a positive control for EGFR blocking and as negative control for oncogene knockdown. Understanding the effect of cetuximab could further explain the underlying mechanism as well as strengthen our earlier hypothesis. DUSP6 is a cytoplasmic gene that plays a pivotal role in the spatiotemporal mechanism of ERK signaling[48]. Our results obtained using RT-qPCR upon oncogenic knockdown of mutant RAS showed downregulation of DUSP6 and the expression of ecto-5′-nucleotidase (NT5E, also known as CD73) remains unchanged. The downregulation of DUSP6 is possibly due to inhibition of ERK based on several negative feedback mechanisms. Indeed, siRNA mediated knockdown affects the ERK pathway leading to the regulation of cytoplasmic genes. However, the effect of EGFR signaling on the gene regulation levels post oncogenic loss need to be monitored. CD73 is a 70 kDa cell surface protein that plays an important role in physiology and pathophysiology of cells. A recent report portrayed the importance of CD73 in relation to EGFR; wherein, it was reported that the CD73 positively regulates EGFR expression levels[34]. In our study, we observed that CD73 expression levels remain unchanged with siRNA knockdown mediated by routine transfection as well as for Ab-Gel$_{GEF}$NP with and without siRNA. The result is in agreement with those obtained through protein regulation studies, since the mere knockdown using siRNA causes no impedance to EGFR which in turn causes no change in cellular mechanism operating under RAS/MAPK pathway, and therefore are independent of EGFR signaling. However, cells transfected with the TBN are rendered susceptible to change in mechanism as well as EGFR dependent signaling causing imbalance of receptor tyrosine kinases. For balancing the requisite EGFR expression, CD73 is overexpressed to compensate for the loss of EGFR caused by the nanoparticle. It is evident that standalone siRNA and cetuximab do not influence CD73 expression levels. Correlating these results, increased level of CD73 activity suggests a synergistic effect of cetuximab and oncogene knockdown. In the case of siRNA transfected cells, albeit there was loss of activity in RAS pathway, a parallel effector pathway governed by EGFR could have been adopted by the cells for survival. However, the nanoparticles experience EGF receptor mediated endocytosis for siRNA delivery, subsequent oncogene knockdown accompanied with loss of activity in the primary effector pathway. The simultaneous disruption of several interdependent pathways and downstream effector proteins, i.e. RAS/MEK/ERK, PI3K/AKT, GAB1 and SHP2, could be the reason for significant change in cellular gene & protein expression levels leading to sensitization towards small molecule tyrosine kinase inhibitor.

## Conclusion

The tri-block nanoparticle platform developed in our study, siRNA conjugated to antibody surface functionalized on gelatin nanoparticles carrying small molecule tyrosine kinase inhibitor, has great potential in siRNA mediated therapy. Interestingly, the platform protects siRNA from degradation and targets the delivery of siRNA to the desired biomarker. Also, the gelatin nanoparticles, used as the carrier system, enable concomitant delivery of drug within the cells of interest. The platform allows loading predetermined and proportional amounts of antibody, siRNA and drug within the carrier for effective targeting and high bioavailability to provide gene and combinational therapy. Using this platform as a therapeutic drug, we found that GAB1 plays a crucial role in the absence of gefitinib for cell survival even after oncogene knockdown. We also found that mutant KRAS oncogene knockdown impairs and alters the localization of SHP2 abrogating its complex formation with GAB1 in the presence of gefitinib leading to the apoptosis of the affected cells. Preliminary *in vivo* safety studies in mice showed TBN is safe to inject with minimal or no toxicity to major organs. It is noteworthy to mention that platform delivery system reported in the work can be modified as per patient needs wherein the siRNA, antibody and the encapsulated drug can be changed with their respective counterparts depending upon the nature and status of cancer.

## Materials and Methods

**Synthesis of Gelatin Nanoparticles.**   Two step desolvation process was used for preparing Gelatin nanoparticles. 500 mg of Gelatin type A (bloom 300) was first dissolved in 10 ml of De-Ionized (DI) water at 50 °C and subjected to first desolvation using rapid addition of acetone (20 ml). The precipitate was dissolved in 10 ml DI water (pH 2.75) and second desolvation using dropwise addition of acetone (3 ml/min) was initiated. Transformation of the transparent solution to milky white solution indicated successful desolvation and formation of nanoparticles. After 10 minutes, the nanoparticles were cross-linked with 200 μl of 25% glutardehyde. The reaction was allowed overnight at 50 °C and the resulting nanoparticulate solution washed 5 times with DI water to remove excess glutaraldehyde (20,000 g for 45 minutes per wash). For the preparation of gefitinib encapsulation, 1 mg of gefitinib hydrochloride dissolved in DI water was added during the second desolvation process prior to acetone addition. The nanoparticles were then resuspended in DI water and stored at 4 °C.

**Cetuximab Conjugation.**   The carboxyl groups present on the surface of the Gel NPs and $Gel_{GEF}$NPs were activated using EDC/NHS reaction. 10 mg of gelatin nanoparticles were suspended in MES buffer (pH 4.5) at a concentration of 5 mg/ml. Activation was performed at room temperature for 3.5 hours under constant shaking (850 RPM). The activated nanoparticles were washed to remove excess EDC/NHS (20,000 g for 20 minutes). The activated particles were resuspended in 2 ml PBS (pH 6.7–7.0) containing 600 ul of cetuximab (2 mg/ml) and the reaction was allowed to incubate overnight at room temperature under constant shaking (850 RPM). The particles were then washed to remove the supernatant containing unreacted antibody. The antibody functionalized particles (Ab-Gel NP) were re-suspended in PBS and stored at 4 °C. Excess antibody through centrifugal separation was analyzed for antibody quantification using Bradford assay with appropriate controls and standards.

**siRNA functionalization.**   Sulfo-SMCC (0.75 mg) was dissolved in 0.5 ml DI water and added to Ab-Gel NPs or Ab-$Gel_{GEF}$NPs (7.5 mg suspended in 1X PBS). The pH of the solution was maintained at 6.7–7.0. The reaction was allowed for 3.5 hours at room temperature and the excess sulfo-SMCC was removed through washing as described previously for EDC/NHS removal. The nanoparticle solution, post conversion of lysine residues on antibody with the SMCC linker, was resuspended in RNAse free water and 0.05 ml of 50 μM thiol modified siRNA was added to the solution for linking the maleimide end of the linker to the thiol end of the siRNA. The reaction was allowed overnight at 4 °C under constant shaking at 500 RPM. The solution was used immediately for further experiments. For quantification of bound siRNA, Cy5 labelled thiol modified siRNA was used. For determining the long-term stability of siRNA within the construct, 5% sucrose (wt. of sucrose/wt. of Gel NP) was added to the TBN and stored at −50 °C.

**Cetuximab quantification.**   Amount of cetuximab conjugated to TBN was determined by quantifying the amount of unconjugated cetuximab present in the supernatant after the conjugation reaction. For example, 20 μl of the supernatant and 20 μl of standards (cetuximab, 2 mg/ml serially diluted till 0.03125 mg/ml) were independently pipetted in 96 well plate. 200 μl Biorad Bradford reagent was added to each of this well and incubated at room temperature for 30 minutes. Intensity of color developed through interaction of the antibody and the reagent was measured using Biotek Cytation 3 spectrophotometer at wavelength of 595 nm and the amount of cetuximab conjugated to the nanoparticles was back calculated and determined.

**Gefitinib Quantification.**   Estimation of amount of gefitinib encapsulated within the TBN was carried out using absorption spectroscopy. 1 ml of the synthesized TBN containing 1 mg/ml of gelatin nanoparticles was completely degraded using 2 mg/ml of protease. The degraded solution was centrifuged at 20,000 g for 30 minutes to ensure no precipitation of particles. The solution was then passed through 10 kDa Amicon filters (10000 g for 10 minutes) the filtrate was characterized for determining the gefitinib content using absorption spectroscopy at 331 nm. Gefitinib standard curve was then used to determine the concentration of the analyzed filtrate.

**siRNA quantification.** Estimation of siRNA present in TBN was determined through florescence spectroscopy. After the addition of 50 μM siRNA tagged with cy5 dye, the reaction was allowed to proceed as detailed above. 1 ml of TBN containing 50 μM of $siRNA_{cy5}$ was then centrifuged at 20,000 g for 20 minutes and the supernatant was analyzed for unbound siRNA. The analysis was carried out in 96 well plate containing 100 μl of sample (triplicate). The excitation and emission of the dye were 646 nm and 660 nm respectively. Biotek cytation 3 spectrophotometer was used for the analysis.

***In vitro* cytotoxicity assay.** H23 human adenocarcinoma non-small cell lung cancer cells (ATCC, USA) and A549 human adenocarcinoma epithelial cells (ATCC, USA) were grown in RPMI 1640 medium supplemented with 4.5 g/L D-glucose, 25 mM HEPES, 0.11 g/L sodium pyruvate, 1.5 g/L sodium bicarbonate, 2 mM L-glutamine, 10% heat-inactivated fetal bovine serum (Altlanta Biologicals, USA) and 0.1% v/v gentamycin. Cells were cultured in a humidified atmosphere of 95% air and 5% $CO_2$ at 37 °C (Thermo Scientific, USA). For determining *in vitro* cytotoxicity, MTT assay was performed by incubating various samples including the nanoparticle of interest on H23 and A549 cells. The concentrations were normalized with respect to gefitinib concentration for all samples (50 μM, 5 μM, 0.5 μM, 0.05 μM, 0.005 μM 0.0005 μM and 0 μM). For samples that did not contain gefitinib, weight of gelatin nanoparticles was used for normalization. Each sample was analyzed in triplicate. After 24 hours of incubation, 10 μl of MTT solution (ATCC, USA) was added and the plate was incubated at 37 °C for 4 hours. Crystals formed were dissolved in 100 μl solubilizing buffer and the plates were kept at 25 °C for 2 hours. The intensity of the color developed after addition of the solubilizing buffer was measured using Biotek Cytation 3 spectrophotometer at 570 nm. Viability of the cells transfected with various samples was then calculated by considering the untreated cells as 100% viable.

**Fluorescence microscopy.** To investigate the cellular uptake of TBN, we incubated H23 cells with various analogues of the nanoparticle. For simplicity, dyes, Fluorescein (fl) and Cy5 (cy5) functionalized to the corresponding ingredient of the nanoparticle is depicted through subscript. The various analogues are as follows: (1) $Gel_{fl}NP$-Ab-$siRNA_{cy5}$, (2) $Gel_{fl}NP$-Ab-siRNA, (3) Gel NP-$Ab_{cy5}$-siRNA, (4) Gel NP-Ab-$siRNA_{cy5}$, (5) $siRNA_{cy5}$ and (6) $Ab_{cy5}$. H23 cells were seeded in 6-well plates ($5 \times 10^5$ cells/well). Cells were grown on a poly-L-lysine treated glass coverslip. 100 μl of each samples (Analogues 1, 2, 3, 4 and 5) were incubated for 4 hours at 37 °C in serum free media. After treatment, resulting cover slips were washed with PBS (1X) to remove unbound particles and microscopic slides were prepared with DAPI nuclear stain. Slides were imaged using a polarized Dark-field fluorescence microscope at 20X magnification.

**Western blot analysis.** Cells were seeded at a density of $1 \times 10^6$ cells/ml and incubated for overnight at 37 °C in 5% $CO_2$ atmosphere. Nanoparticle and relevant control samples were incubated in serum free media for the period of 72 hrs. For control experiments, siRNA transfection (240 nM) was performed using TransIT-X2 dynamic delivery system transfecting agent (Mirus Bio) as per manufacturer's instructions. Whole-cell lysates were prepared using Triton X 100 lysis buffer with MS-SAFE protease and phosphatase cocktail inhibitor (Sigma-Aldrich) and the protein concentration was equalized by Bicinchoninic acid assay (Sigma-Aldrich). Proteins were separated by 4–15% native PAGE (Bio-Rad) and were transferred onto nitrocellulose membranes (GenScript). Membranes were incubated with primary antibody overnight at 4 °C, were washed and incubated with secondary antibody. Primary antibodies used for western blotting are rabbit polyclonal anti-β-actin, rabbit monoclonal anti-AKT, rabbit monoclonal anti-phospho-AKT, rabbit polyclonal anti-MEK1/2, rabbit polyclonal anti-phospho-MEK1/2, rabbit monoclonal anti-EGFR, rabbit monoclonal anti-phospho-EGFR, rabbit monoclonal anti-SHP2, rabbit polyclonal anti-phospho-SHP2 (Tyr542), rabbit polyclonal anti-phospho-SHP2 (Tyr580), rabbit monoclonal anti-phospho-GAB1 (Tyr627), rabbit polyclonal anti-phospho-GAB1 (Tyr307), all from Cell Signaling and mouse monoclonal anti-KRAS from Santa Cruz Biotech. The membranes were developed with peroxidase-labeled anti-mouse or anti-rabbit IgG (Cell Signaling Tech.) using enhanced chemiluminescence substrate (Pierce) and imaged on Fujifilm LAS-3000 imaging system. Actin protein levels were used as a control for adequacy of equal protein loading. Protein expression levels were quantified by densitometry analysis.

***In vitro* stability studies.** An aliquot of each sample containing 75 pmoles of siRNA was analyzed by 4–15% polyacrylamide gel electrophoresis (PAGE, BioRad, 100 V for 60 minutes) using Tris borate EDTA running buffer. The gels were then stained with GelRed (Thermo Scientific, USA) and imaged using UV trans-illumination light filter and image analysis was performed using Biorad Laboratories Image Lab v.3.0.

**Animal studies.** All animal experiments were approved by the Notre Dame IACUC (approval #14-04-1726) and conducted in accordance with Freimann Life Science Center guidelines for human animal treatment. *In vivo* dose safety studies of TBN were performed in normal mice. TBN (80 mg/Kg of body weight) was suspended in PBS and administered via tail vein for 3 consecutive days. Animals were monitored for any changes in the behavior. After 3 days, animals were sacrificed and vital organs collected, tissues fixed in formalin, stained with H&E for histology.

### Acknowledgements

R.K. kindly acknowledges the "Michael J and Sharon R Bukstein Chair in Cancer Research" for financial support. Authors also acknowledge financial support from Ellis Fischel Cancer Center gift funds, UM-Fast Track Funding, Mizzou Advantage Funding, and Coulter Translational Partnership Grants.

### Author Contributions

R.K. hypothesized the study and designed all the experiments. R.S. and A.Z. performed synthesis and characterization of Gelatin-Ab-siRNA nanoparticles. D.S. and A.Z. performed cell experiments including Western blots and MTT assay. K.T. performed RT-PCR on cell samples. S.C., M.L. and A.U. designed and performed *in vivo* studies. All authors contributed equally in analyzing the data and writing the manuscript.

### Additional Information

**Supplementary information** accompanies this paper at http://www.nature.com/srep

**Competing financial interests:** The authors declare no competing financial interests.

**How to cite this article**: Srikar, R. *et al.* Targeted nanoconjugate co-delivering siRNA and tyrosine kinase inhibitor to KRAS mutant NSCLC dissociates GAB1-SHP2 post oncogene knockdown. *Sci. Rep.* **6**, 30245; doi: 10.1038/srep30245 (2016).